# Efficient production of nuclear isomer $^{93m}$Mo with laser-accelerated proton beam and an astrophysical implication on $^{92}$Mo production

Wenru Fan[1] • Wei Qi[2] • Jingli Zhang[1] • Zongwei Cao[1] • Haoyang Lan[3] • Xinxiang Li[1] • Yi Xu[4] • Yuqiu Gu[2] • Zhigang Deng[2] • Zhimeng Zhang[2] • Changxiang Tan[1] • Wen Luo[1]* • Yun Yuan[1]§ • Weimin Zhou[2]#

[1]School of Nuclear Science and Technology, University of South China, 421001, Hengyang, China

[2]Science and technology on Plasma Physics Laboratory, Laser Fusion Research Center, China Academy of Engineering Physics, 621900, Mianyang, China

[3]State Key Laboratory of Nuclear Physics and Technology, and Key Laboratory of HEDP of the Ministry of Education, CAPT, Peking University, 100871, Beijing, China

[4]Extreme Light Infrastructure - Nuclear Physics (ELI-NP), Horia Hulubei National Institute for R&D in Physics and Nuclear Engineering (IFIN-HH), 30 Reactorului Str., 077125 Buchurest-Magurele, Romania

Wei Qi: Contributing first author; *Corresponding authors: wen.luo@usc.edu.cn; §Corresponding authors: yuanyun_usc@qq.com; #Corresponding authors: zhouwm@caep.cn

**Abstract:** Nuclear isomers play a key role in the creation of the elements in the universe and have a number of fascinating potential applications related to the controlled release of nuclear energy on demand. Particularly, $^{93m}$Mo isomer is a good candidate for studying the depletion of nuclear isomer via nuclear excitation by electron capture. For such purposes, efficient approach for $^{93m}$Mo production needs to be explored. In the present work, we demonstrate experimentally an efficient production of $^{93m}$Mo through $^{93}$Nb(p, n) reaction induced by intense laser pulse. When a ps-duration, 100-J laser pulse is employed, the $^{93m}$Mo isomer at 2425 keV (21/2$^+$, $T_{1/2}$ = 6.85 h) are generated with a high yield of 1.8 × 10$^6$ particles/shot. The resulting peak efficiency is expected to be 10$^{17}$ particles/s, which is at least five orders of magnitudes higher than using classical proton accelerator. The effects of production and destruction of $^{93m}$Mo on the controversial astrophysical p-isotope $^{92}$Mo are studied. It is found that the $^{93}$Nb(p, n)$^{93m}$Mo reaction is an important production path for $^{93m}$Mo seed nucleus, and the influence of $^{93m}$Mo-$^{92}$Mo reaction flow on $^{92}$Mo production cannot be ignored. In addition, we propose to directly measure the astrophysical rate of (p, n) reaction using laser-induced proton beam since the latter one fits the Maxwell-Boltzmann distribution well. We conclude that laser-induced proton beam opens a new path to produce nuclear isomers with high peak efficiency towards the understanding of p-nuclei nucleosythesis.

**Introduction**

Nuclear isomers are relatively long-lived 'metastable' excited states, with half-lives ranging from nanoseconds to years. The long half-lives of such isomers provide a unique opportunity for exploring nucleosynthesis in extreme astrophysical surroundings[1,2]. When carrying out an astrophysical nucleosynthesis network calculation, highly accurate inputs of nuclear reaction rates are expected, and even one reaction rate can dramatically influence the whole astrophysical evolution network[3,4]. Nuclear isomers also have a broad range of potential applications including nuclear batteries[5-7], medical isotopes[8-10], nuclear clocks[11,12], and nuclear gamma-ray lasers[13]. For example, nuclear battery composed of nuclear isomers have a much higher energy density than chemical batteries and hence is considered as a good energy source candidate for deep space exploration.

Recently, the manipulation of nuclear isomer $^{93m}$Mo has gained lots of attention. $^{93m}$Mo has a (21/2)$^+$ isomer at 2,425 keV with a half-life of 6.85 h and a (17/2)$^+$ intermediate state that lies 4.85 keV higher at 2,430 keV with a half-life of 3.5 ns. Such isomeric property is attractive to nuclear medical applications[14] and to exploiting the depletion of nuclear isomers via nuclear excitation by electron capture (NEEC)[15-18]. Chiara *et al.* provided an experimental evidence of $^{93m}$Mo isomer depletion caused by NEEC. However, the reported depletion probability is a few orders of magnitudes higher than the theoretical expectation[19] and such isomer depletion can not be confirmed by another dedicated experiment[20]. In order to investigate the depletion of nuclear isomer $^{93m}$Mo, it is a necessity to efficiently produce them.

One of the greatest questions for modern physics to address is how elements heavier than iron are created in extreme astrophysical environments. A particularly challenging part of that question is the creation of the so-called 35 p-nuclei between $^{74}$Se and $^{196}$Hg, which can be produced under explosive conditions in a sequence of photodisintegration reactions[21]. Among these p-nuclei, $^{92}$Mo (neutron number N = 50) is a two neutron-magic p-isotope and exhibits larger abundances than the neighboring p-nuclei. Fig. 1 shows the production of $^{92}$Mo isotope by different photodisintegration paths. These paths include $^{93}$Mo(γ, n), $^{93}$Tc(γ, p) and $^{96}$Ru(γ, α). Recently, a nucleosynthesis sensitivity study shows that the $^{92}$Mo isotope is mainly produced by photodisintegration of $^{93}$Mo[22]. However, the production of $^{92}$Mo in stars is a puzzle for nuclear astrophysics since present models underestimate this p-isotope by a few orders of magnitudes. When calculating astrophysical rates, the isotopes are typically assumed to be in their ground state, or their levels are populated according to the thermal equilibrium probability distribution[4]. Those assumptions may not be valid for actual astrophysical circumstances. Some nuclear isomers, which play an essential role in nucleosynthesis, should be treated as special isotopes, i.e., astromer, as is called in Ref.[23]. Does the "special isotope" $^{93m}$Mo affect the abundance of $^{92}$Mo via the subsequent (γ, n) reaction (see Fig. 1)? It is worth considering whether the $^{93m}$Mo-$^{92}$Mo reaction flow is also involved in the production of the missing p-isotope $^{92}$Mo.

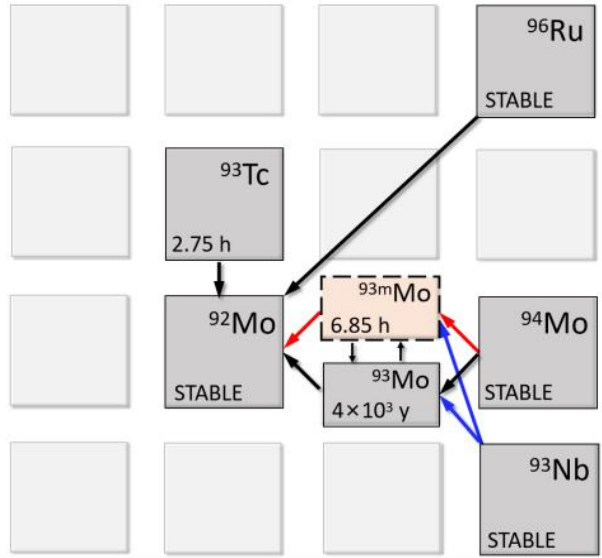

**Fig. 1.** An excerpt of the chart of nuclei depicting the production of $^{92}$Mo via photodisintegration reactions. The red lines indicate the reaction flow through $^{93m}$Mo that may affect the abundance of $^{92}$Mo in extreme astrophysical environments. The blue lines show the production of $^{93}$Mo in the ground state and isomeric state via proton induced reactions.

The production of $^{93m}$Mo has been realized by using radio-frequency accelerators[24-27]. In the recent experiment showing experimental evidence of NEEC, with a reported beam intensity of about $6 \times 10^8$ ions/s, the total production rate of $^{93m}$Mo is estimated to be about 9.3 kHz[28]. It should be noted that such production rate is not efficient for the following isomer depletion, especially when the NEEC rate is not sufficiently high. With the rapid development of ultraintense and ultrashort laser technology[29], laser intensity focused on the target can exceed $10^{22}$ W/cm$^2$. High-intensity lasers can produce energetic particle beams with large charge for many pioneering applications[30-33]. It is demonstrated that $^{115m}$In isomer was produced by laser-accelerated electron beam, through (γ, xn) reaction, with a peak efficiency of $1.76 \times 10^{15}$ particles/s[34]. $^{83m}$Kr isomer was populated successfully by Coulomb collision of ions with quivering electrons with a peak efficiency of $2.34 \times 10^{15}$ particles/s, which they stated is five orders of magnitudes higher than the one achieved by classical accelerator[35].

In this work, we demonstrate experimentally an efficient production of $^{93m}$Mo by laser-accelerated proton beam, through $^{93}$Nb(p, n) reaction, as schematically shown in Fig. 2. When a ps-duration, 100-J laser pulse is employed, the $^{93m}$Mo isomer at 2425 keV ($21/2^+$, $T_{1/2}$ = 6.85 h) are generated with a high yield of $1.8 \times 10^6$ particles/shot. The resulting peak efficiency exceeds $10^{17}$ particles/s. The effect of nuclear reaction flow on the population of $^{93m}$Mo is studied. The $^{93m}$Mo involved photodisintegration reactions leading to the production of $^{92}$Mo, which is one of the most debated p-nuclei, is then discussed. Finally, the astrophysical reaction rate of $^{93}$Nb(p, n)$^{93m}$Mo at the p-process temperatures is directly deduced with laser-accelerated proton beam since the latter one matches well with the Maxwell-Boltzmann (M-B) distribution at high energies[36].

**Results**

**Experimental setup.** The laser proton acceleration (LPA) and the following $^{93m}$Mo population experiments were performed with the XingGuangIII laser facility at the Laser Fusion Research Center (LFRC) in Mianyang. The experimental setup is schematically shown in Fig. 2(a). In the first stage of our experiments, a laser pulse of 840 fs (FWHM), with 116 J of energy is focused by an off-axis parabolic mirror onto a thin Cu foil with thickness varying from 7 to 15 µm, generating a high-energy proton beam which enables triggering the $^{93}$Nb(p, n)$^{93m}$Mo reaction. A Radiochromic Film (RCF) stack is used to diagnose the angular distribution of the proton beam and a Thomson Parabolas Spectrometer (TPS) to detect the energy spectrum of protons passing through a hole in the center of the RCF stack. In the second stage, the RCF stack and the TPS are removed, and the laser-accelerated proton beam impinges directly the Nb target, which has a natural abundance of 99.9% and thickness of 1 mm and is placed ~14 mm downstream from the Cu foil. The $^{93}$Mo isotope in ground and excited states are then produced through (p, n) reactions. After irradiation, the Nb target is taken out from the target chamber of the XingGuangIII laser facility. An off-line detection of characteristic γ emissions of Nb target is then performed with a $^{60}$Co calibrated high-purity germanium (HPGe) detector.

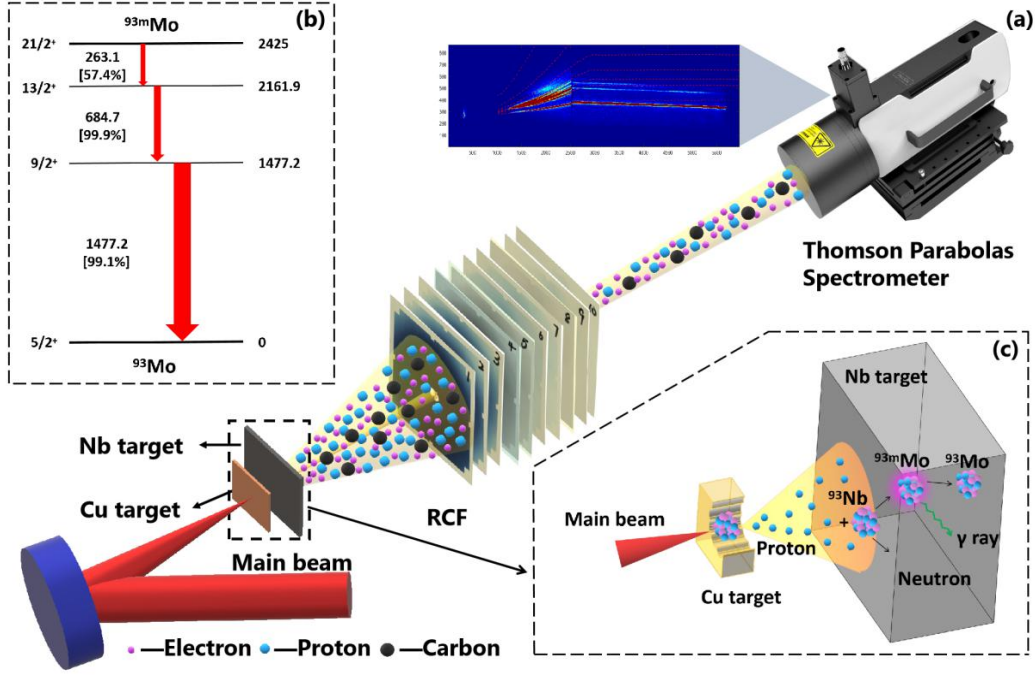

**Fig. 2. (a)** Schematic diagram of the experimental setup for nuclear isomer $^{93m}$Mo production at the XingGuangIII laser facility (not to scale). **(b)** Partial level scheme (not to scale) for the $^{93m}$Mo nucleus ($Z$ = 42). The right side of the panel gives the level energies (in keV) and half-lives, and the left side gives the angular momenta and parities. **(c)** Schematic diagram of the target arrangement for the LPA and the following $^{93m}$Mo production.

## A. Laser accelerated proton beam

Recently, laser-accelerated proton beams within the MeV range have drawn considerable interest because of their potential applications in experimental astrophysics[37], medical isotope production[38,39], inertial confinement fusion[40]. In our experiments, the energetic protons used for $^{93m}$Mo population are mainly produced through the target normal sheath acceleration (TNSA)[41]. The LPA is optimized by varying the thickness of Cu foil. For three thicknesses of 7, 10 and 15 μm, the energy spectra and angular distributions of the laser-accelerated proton beam are displayed in Fig. 3. It is shown that the proton energy $E_p$ exceeds 10 MeV and the energy spectra fit well with the following M-B distribution[42]:

$$f(E) = \frac{E}{(kT)^2} exp\left(-\frac{E}{kT}\right), \quad (1)$$

where $k$ is the Boltzmann constant and $T$ is the proton temperature. In the 7-μm-thick and 10-μm-thick cases, the measured proton spectra indicates two-temperature M-B distributions. These temperatures are fitted to be $T_1$ = 1.14 MeV, $T_2$ = 2.95 MeV for the 7-μm Cu foil, and $T_1$ = 1.29 MeV, $T_2$ = 2.76 MeV for the 10-μm Cu foil. However, a single M-B distribution with $T$ = 1.51 MeV appears in the case of 15-μm Cu foil. The total charge of protons with energies above 4 MeV, which is slightly higher than the reaction threshold of $^{93}$Nb(p, n)$^{93m}$Mo (= 3.7 MeV), are obtained to be 5.05 × 10$^{12}$ particles/shot (808 nC), 5.10 × 10$^{12}$ particles/shot (816 nC) and 2.52 × 10$^{12}$ particles/shot (403 nC) for 7-μm-thick, 10-μm-thick and 15-μm-thick Cu foil, respectively. Fig. 3(b) demonstrates a strong correlation between proton energy and angular divergence. It decreases with increasing proton energy. The maximum divergence angle is about 15 degrees, which is independent to the thickness of Cu foil in our case. Taking into account the proton charge and temperature, the 10-m-thick Cu foils were selected for the subsequent $^{93m}$Mo population experiments.

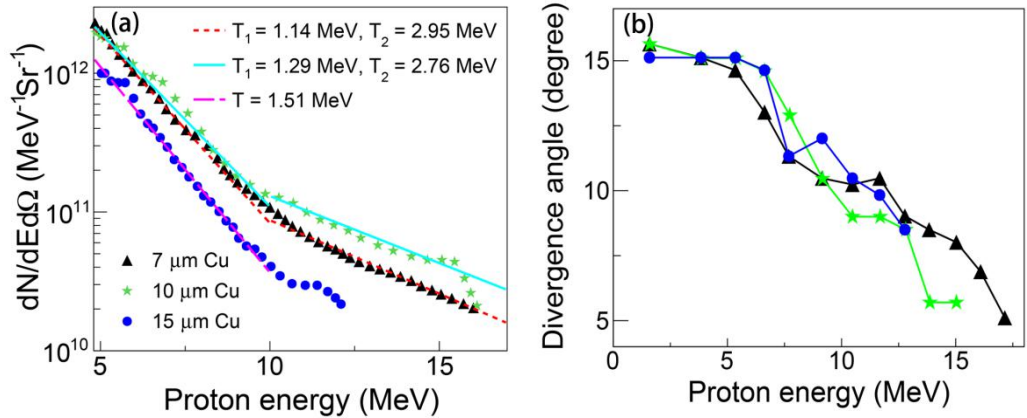

**Fig. 3.** The energy spectra **(a)** and angular patterns **(b)** of protons obtained by irradiating Cu foils with ps-duration, 100-J laser pulses.

B. **$^{93m}$Mo isomer production**

The Nb target used for $^{93m}$Mo production was placed inside the vacuum chamber. After irradiation, it took approximately 30 minutes to reduce the vacuum level of the chamber before the target could be removed for off-line detection. The produced $^{93m}$Mo isomer has a half-life of 6.85 h and three principal characteristic emissions at energies of 263.05, 684.69, and 1477.14 keV. Their respective branching intensity are

$I_\gamma$ = 57.4%, 99.9%, and 99.1%. These characteristic γ rays emitted from the Nb targets were detected successfully with the HPGe detector, as shown in Fig. 4 (a). During the off-line detection, the Nb target was positioned 1.4 cm away from the endcap of the detector, which weakens the sum-peak effect induced by simultaneous detection of two or more characteristic γ rays, as discussed later. Fig. 4(b) displays the cumulative peak counts of the three characteristic γ-ray lines as a function of measurement time. The fitting function follows the typical formula for nuclear decay:

$$N_{det} = N_0 \left\{1 - \exp\left(-\frac{ln2}{T_{1/2}}t\right)\right\}, \qquad (2)$$

where $N_{det}$ is the peak count accumulated at time instant $t$, $N_0$ is the total peak count when all $^{93m}$Mo isomers finish their decays. According to Eq. (2), the half-lives of the three γ-ray lines at energies of 263.05, 684.69, and 1477.14 keV are determined to be 6.89 ± 0.06, 6.95 ± 0.10, and 6.81 ± 0.10 h, respectively. These results are in good agreement with the theoretical value of $^{93m}$Mo ($T_{1/2}$ = 6.85 h) provided by the NNDC database[43].

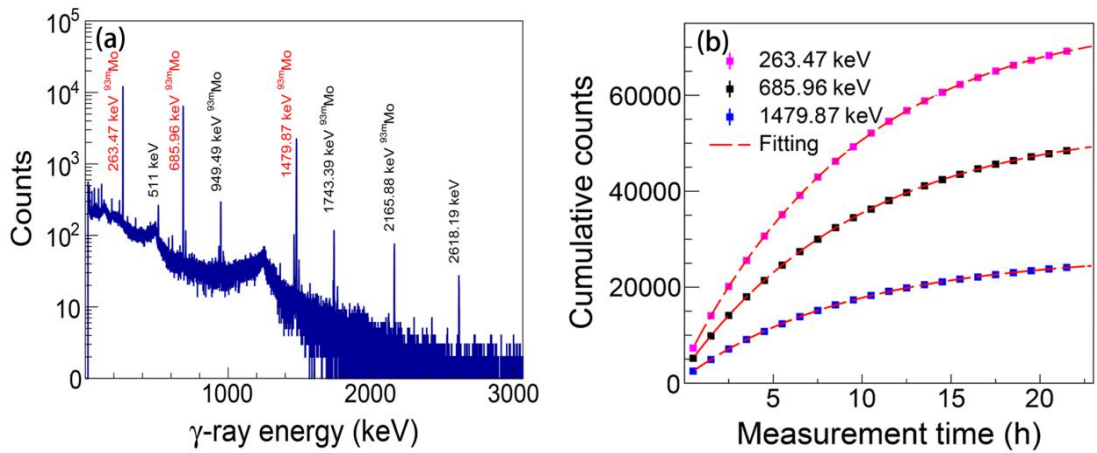

**Fig. 4.** (a) Examplary HPGe γ ray spectrum for Nb target with a measurement time of 22 h. **(b)** Peak counts accumulated for three characteristic γ-ray lines at energies of 263.47, 685.96, and 1479.87 keV.

As mentioned above, the laser-accelerated proton beam follows the M-B distribution. Using Eq. (1). the production yield $Y_{exp}$ of $^{93m}$Mo isomer can be written as:

$$Y = \frac{nd}{(kT)^2} N_p \int_0^\infty \sigma(E) E \exp\left(-\frac{E}{kT}\right) dE, \quad (3)$$

where $n$ denotes the number density of the target nuclei, $d$ represents the thickness of the Nb target, $N_p$ is the total number of energetic protons irradiating the Nb target, and $\sigma(E)$ represents the cross section for $^{93m}$Mo through proton induced reaction. Fig. 5 shows the experimental and calculated $\sigma(E)$ for $^{93}$Nb(p, n)$^{93m}$Mo reaction. The calculated data exhibit a satisfactory agreement with the available experimental ones for proton energies lower than 16 MeV. The reaction cross section is peaked at $E_p$ = 12.5 MeV and the maximum cross section is ~35 mb.

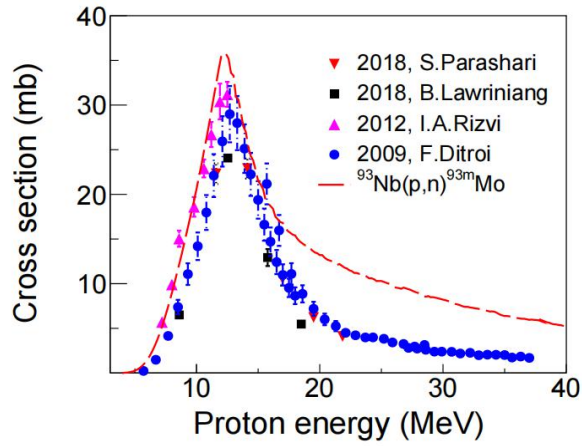

**Fig. 5.** Cross sections curve of $^{93}$Nb(p, n)$^{93m}$Mo reaction calculated by Talys software[44] together with the available experimental data[45-48].

The yield of $^{93m}$Mo isomers produced in the experiment can be obtained by the peak count of characteristic γ rays. Its expression can be written as:

$$Y_{exp} = \frac{N_{det}}{PI_\gamma \varepsilon \{1-\exp(-\lambda t_m)\}\exp(-\lambda t_d)}, \quad (4)$$

where $t_m$ is the real measurement time, $t_d$ is the delay (cooling) time between target irradiation and initial detection, $\lambda$ represents the decay constant of the $^{93m}$Mo isomer, and ε represents the detection efficiency of the HPGe detector. At $t_m$ = 22 h, the $N_{det}$ values for the three characteristic γ rays at 263.05, 684.69, and 1477.14 keV are (6.92 ± 0.03) × 10$^4$, (4.84 ± 0.03) × 10$^4$, and (2.43 ± 0.02) × 10$^4$ per shot, respectively. According to Eq. (4), the production yield $Y_{exp}$ are obtainted to be (1.77 ± 0.01) × 10$^6$, (1.66 ± 0.01) × 10$^6$ and (1.60 ± 0.01) × 10$^6$ per shot, respectively. Table 1 summarizes the basic parameters of three principal characteristic γ-ray lines and the resulting production yield for $^{93m}$Mo isomers. There are visible differences between the production yields of $^{93m}$Mo isomer that obtained by the peak counts of three characteristic γ rays. Such discrepancy can be attributed to the sum-peak effect. As illustrated in Fig. 4(a), apart from the three primary characteristic γ-ray peaks, three additional high-energy γ rays are located at 949.49, 1743.39, and 2165.88 keV. Through fitting the curves for their peak counts, their respective half-lives are determined to be 6.89 ± 0.06, 6.95 ± 0.10, and 6.81 ± 0.10 h, respectively, which are consistent with the recommended half-life ($T_{1/2}$ = 6.85 h) of the $^{93m}$Mo isomer within the statistical uncertainty. This indicates that these peaks also originate from the radiation decay of $^{93m}$Mo isomer at 2,425 keV. For example, the 949.49 keV peak is caused by the sum of two characteristic γ-ray lines at 263.1 and 684.7 keV. Considering the sum-peak effect, the $Y_{exp}$ value is corrected to be ~1.8 × 10$^6$ per shot.

Table 1. The basic parameters of three principal characteristic γ-ray lines and the resulting production yield for $^{93m}$Mo isomers.

| Isomer | $E_\gamma$ (keV) | $I_\gamma$ (%) | $E_{\gamma, exp}$ (keV) | $T_{1/2, exp}$ (h) | $N_{det}$ (×10$^4$) | $Y_{exp}$ (×10$^6$) |
|---|---|---|---|---|---|---|
| $^{93m}$Mo | 263.05 | 57.4% | 263.47 ± 2.62 | 6.89 ± 0.06 | 6.92 ± 0.03 | 1.77 ± 0.01 |
| | 684.69 | 99.9% | 685.96 ± 1.65 | 6.95 ± 0.10 | 4.84 ± 0.03 | 1.66 ± 0.01 |
| | 1477.14 | 99.1% | 1479.87 ± 4.31 | 6.81 ± 0.10 | 2.43 ± 0.02 | 1.60 ± 0.01 |

Fig. 6(a) depicts the simulated temporal evolution of $^{93m}$Mo isomer considering laser-accelerated proton beams generated by the different thicknesses of Cu foils, as shown in Fig. 3. For 7-μm and 10-μm thick Cu foil, the production duration of $^{93m}$Mo isomers is estimated to be 65 ps. As a result, the peak efficiency of $^{93m}$Mo is evaluated to be 2.77 × 10$^{16}$ particles/s. The spatial distribution of $^{93m}$Mo produced by $^{93}$Nb(p, n) reaction is shown in Fig. 6(b). One can see that the $^{93m}$Mo isomers are predominantly generated in the central region of the transverse plane (X-Y plane), and the isomer density decreases with increasing transversal distance. This is attributed to the fact that high-energy protons with smaller divergence angles have a greater probability to produce $^{93m}$Mo isomers. Along the longitudinal direction (Z-axis), the isomer density decreases as a result of limited penetration depth of the proton beam. Note that this trend is beneficial to optimizing the target dimension employed for isomer or isotope production[49].

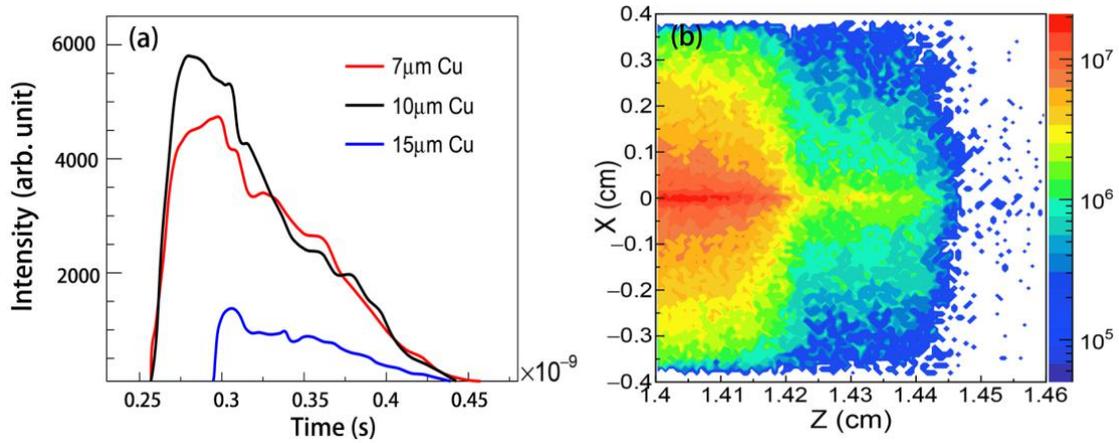

**Fig. 6.** The simulated temporal **(a)** and spatial **(b)** distributions of $^{93m}$Mo isomers produced in the Nb target. The experimental data for the proton beam shown in Fig. 3 are used in the simulations. To expedite the computational process and improve the statistical precision, the simulated cross section data of $^{93}$Nb(p, n) reaction shown in Fig. 5 are artificially amplified by 100 times in the simulation.

It is very interesting to investigate the dependence of $^{93m}$Mo excitation on the proton temperature. It is shown in Fig. 7(a) and 7(b) that the production yield of $^{93m}$Mo increases with increasing proton temperature, whereas the production duration keeps almost unchanged. Accordingly, the peak excitation efficiency of $^{93m}$Mo is enhanced with the proton temperature, however, the growth becomes slow down at relatively high proton temperature, as shown in Fig. 7 (b). When proton temperature becomes higher than 2.25 MeV, the peak excitation efficiency exceeds $10^{17}$ particle/s. Such efficient isomer excitation may be helpful for exploring NEEC and NEET effects since currently the detection of NEEC and NEET processes is very challenging due to extremely low isomer depletion probability and significant background signals[50].

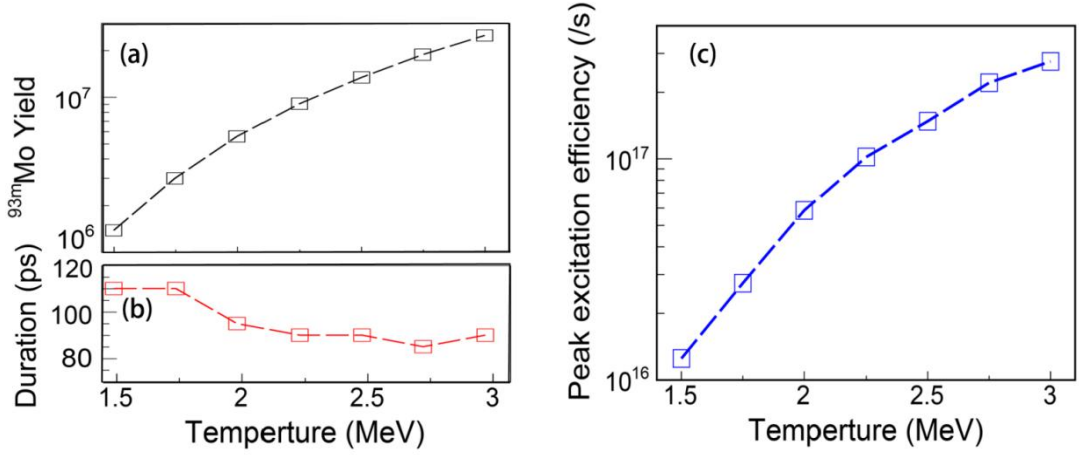

**Fig. 7.** The $^{93m}$Mo yield **(a)**, duration **(b)** and the resulting peak excitation efficiency **(c)** as a function of proton temperature. The charge of proton beam used in the simulations are 816 nC.

## C. Astrophysical implication on $^{92}$Mo production

As aforementioned, the underestimation of abundance of the p-nucleus $^{92}$Mo in stars is a pending problem in nuclear astrophysics, which may result from the lack of consideration for isomer contributions in nuclear network calculations. The astrophysical reaction rate of (γ, n) can be written by[51]:

$$r_\gamma = \frac{8\pi}{h^3 c^2} \int_0^\infty \frac{E_\gamma^2}{exp(E_\gamma/KT)-1} \sigma(E_\gamma) dE_\gamma, \qquad (5)$$

where $h$ is the Planck constant, $c$ is the velocity of light in vacuum, and $E_\gamma$ is the γ-photon energy. To briefly evaluate the effect of $^{93m}$Mo destruction on $^{92}$Mo production, the cross section of $^{93m}$Mo(γ, n)$^{92}$Mo reaction and its astrophysical reaction rate are calculated and then compared with those of $^{96}$Ru(γ, α), $^{93}$Tc(γ, p) and $^{93}$Mo(γ, n) reactions, as shown in Fig. 8. At the p-process temperatures, the cross section curves of four photodisintegration reactions leading to the $^{92}$Mo production are peaked at almost the same proton energy, ~16 MeV.. The maximum cross sections for $^{93m}$Mo(γ, n) and $^{93}$Mo(γ, n) reactions are higher than those for $^{96}$Ru(γ, α), $^{93}$Tc(γ, p) reactions. Fig. 8(b) shows that although the astrophysical rate of $^{93m}$Mo(γ, n)$^{92}$Mo is visibly lower than the $^{93}$Tc(γ, p)$^{92}$Mo since the latter one has a significant lower reaction threshold, however, it is still comparable to those of $^{93}$Mo(γ, n) and $^{96}$Ru(γ, α) reactions. This suggests that the $^{93m}$Mo(γ, n)$^{92}$Mo should be considered as an alternative nuclear reaction flow for $^{92}$Mo production at high temperatures of astrophysical interest.

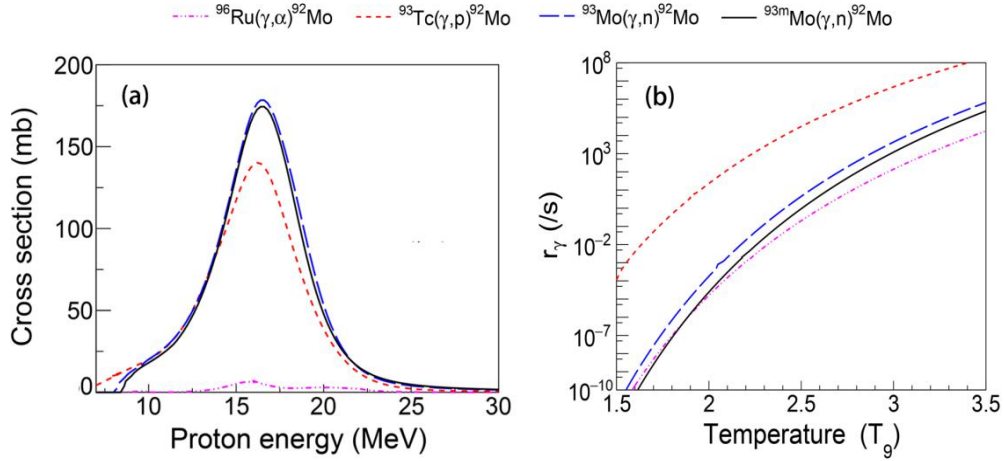

**Fig. 8. (a)** Cross section curves of $^{96}$Ru(γ, α), $^{93}$Tc(γ, p), $^{93}$Mo(γ, n) and $^{93m}$Mo(γ, n) reactions leading to $^{92}$Mo production. **(b)** The astrophysical rate of $^{96}$Ru(γ, α), $^{93}$Tc(γ, p), $^{93}$Mo(γ, n) and $^{93m}$Mo(γ, n) as a function of astrophysical temperature $T_9$.

The seed nucleus $^{93m}$Mo can be synthesized from $^{94}$Mo(γ, n), $^{95}$Mo(γ, 2n) and $^{93}$Nb(p, n) reactions. Fig. 9(a) shows the spectral distributions of proton and photon flows at the p-process relevant temperature of $T_9$ = 3.0. For both the proton and photon flows, their intensities show a flat distribution within energy range of 0.1-1.0 MeV and then decrease rapidly at higher energies. The comparison between cross section curves of $^{94}$Mo(γ, n), $^{95}$Mo(γ, 2n) and $^{93}$Nb(p, n) reactions are further presented in Fig. 9(b). The maximum cross section of $^{93}$Nb(p, n)$^{93m}$Mo is more than one order of magnitude higher than the other two reaction paths. Meanwhile, the $^{93}$Nb(p, n)$^{93m}$Mo reaction has a visibly lower reaction threshold. Hence, low-energy protons can be sufficiently involved in the astrophysical nucleosynthesis of $^{93m}$Mo via proton induced reaction, which may result in a large astrophysical rate. It is expected that the $^{93}$Nb(p, n) reaction could play an important role for production of seed nucleus $^{93m}$Mo, whose destruction by (γ, n) reaction is non-negligible for the population of the debated p-nucleus $^{92}$Mo.

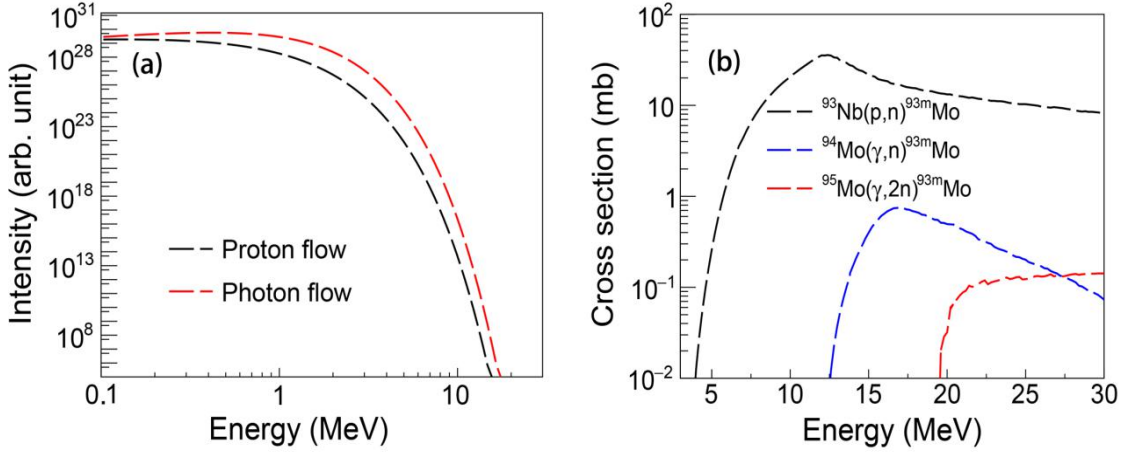

**Fig. 9. (a)** The spectral patterns of protons and photon flows at the astrophysical temperature of T$_9$ = 3.0. The proton pattern is calculated with Eq. (1) and the photon pattern is considered as the Planck spectrum. In order to realize the same level for particle flow intensity at energy of 0.1 MeV, the proton flow intensity is enhanced artificially by a factor of 1.13 × 10$^{35}$. **(b)** The cross section curves of $^{93}$Nb(p, n), $^{94}$Mo(γ, n) and $^{95}$Mo(γ, 2n) reactions leading to $^{93m}$Mo production.

**Discussion**

In nuclear astrophysics, the astrophysical rate is generally defined as the number of nuclear reactions occurring per unit time[52]. In this study, we limit it to the two-body interaction between two nuclei. Since the spectral distributions of the particle flow in astrophysical plasmas can be approximated to the M-B function, the proton-induced astrophysical rate follows[53]:

$$r_p = \sqrt{\frac{8}{\pi\mu}}\left(\frac{1}{KT}\right)^{3/2} \int_0^\infty \sigma(E) E \exp\left(-\frac{E}{kT}\right) dE, \qquad (6)$$

where $T_p$ symbolizes the plasma temperature, σ(E) denotes the reaction cross section, and μ is defined as $A_1 A_2/(A_1 + A_2)$ with $A_1$ and $A_2$ being the mass numbers of the two nuclei involved in the two-body reaction. In our cases, the laser-accelerated proton spectrum resulted from TNSA follows the M-B distribution, which are in agreement with the one predicted in astrophysical environments. Substituting Eq. (3) into Eq. (6), the astrophysical rate $r_p$, can be directly correlated to the production yield Y:

$$r_p = \sqrt{\frac{8KT}{\pi\mu}}\frac{1}{ndN_p}Y. \tag{7}$$

This indicates that the astrophysical rate can be obtained by the energy spectrum of the laser-accelerated proton beam, if only such spectral pattern matches well with the single M-B distribution. It is worth noting that in the current experiment, the astrophysical rates for p-process nucleosynethesis can still not be derived. This is because the laser-accelerated proton beam is constituted of M-B distribution with two temperatures, as discussed above. These temperatures are higher than the typical temperature for p-process nucleosynethesis (0.22-0.30 MeV). As a consequence, it is expected that additional laser-plasma experiments are performed to obtain a proton beam with single M-B distribution at temperatures of astrophysical interest.

**Conclusion**

We have demonstrated experimentally that $^{93m}$Mo isomers are efficiently populated by laser-accelerated proton beam with large charge, through (p, n) reaction. The production yield of $^{93m}$Mo reaches $1.8 \times 10^6$ particles/shot, and the resulting peak excitation efficiency is expected to be more than $10^{17}$ particles/s, which is at least five orders of magnitudes higher than using traditional accelerator. This highly efficient excitation of nuclear isomers within the picosecond duration is very useful for the study of isomer depletion processes including NEET and NEEC. We further propose to directly reveal astrophysical rates by using laser-accelerated proton beam, given that the latter one conforms closely to the M-B distribution. Moreover, the effects of population and destruction of $^{93m}$Mo isomer on the debated p-isotope $^{92}$Mo are studied. It is found that influence of the reaction flow from $^{93m}$Mo to $^{92}$Mo cannot be ignored, and the $^{93}$Nb(p, n)$^{93m}$Mo reaction is an alternative route for production of the seed nucleus $^{93m}$Mo. We conclude that laser-induced proton beam could open a new path to produce nuclear isomers with high peak efficiency towards understanding the origin of p-nuclei

**Methods**

**Geant4 simulation.** To model the interaction of laser-accelerated proton beam with Nb target, we employ the Geant4-GENBOD toolkit[54,55], which uses as input photonuclear cross-section data from theoretically computed or experimental databases. This toolkit has been used for studies of medical isotope production using intense laser-plasma electron sources[49]. The experimental spectral and angular distributions of proton beam (see Fig. 3) and the simulated cross section curve for $^{93}$Nb(p, n) reaction (see Fig. 4) were employed in the Geant4 simulations. The Nb target used for proton irradiation has ~100% natural abundance. The target geometry and installation also followed the experimental setup.

**Talys calculation.** The calculations of the cross sections and the astrophysical reaction rates were performed with Talys 1.9[44], which generate the results illustrated in Fig. 5, Fig. 8 and Fig. 9(b). The nuclear structure ingredients used for the Talys computations are explicitly presented in Lan Haoyang's article[56].

**ACKNOWLEDGMENTS**

We would like to express our thanks to the XingGuang III operation team for their great supports. This work was supported by the National Key R&D Program of China (Grant No. 2022YFA1603300), the National Natural Science Foundation of China (Grant No. U2230133), the Independent Research Project of the Key Laboratory of Plasma Physics, CAEP (Grant No. JCKYS2021212009), the Open Fund of the Key Laboratory of Nuclear Data, CIAE (Grant No. JCKY2022201C152), the Research Foundation of Education Bureau of Hunan Province, China (22B0453) and Hengyang

**Competing interests**

The authors declare no competing interest.